# ON THE MAPPING OF TIME-DEPENDENT DENSITIES ONTO POTENTIALS IN QUANTUM MECHANICS

Roi Baer

*Department of Physical Chemistry and the Fritz Haber Center for Molecular Dynamics, the Hebrew University of Jerusalem, Jerusalem 91904 Israel.*

Draft: Friday, April 27, 2007

The mapping of time-dependent densities on potentials in quantum mechanics is critically examined. The issue is of significance ever since Runge and Gross (Phys. Rev. Lett. 52, 997 (1984)) established the uniqueness of the mapping, forming a theoretical basis for time-dependent density functional theory. We argue that besides existence (so called v-representability) and uniqueness there is an important question of stability and chaos. Studying a 2-level system we find innocent, almost constant densities that cannot be constructed from any potential (non-existence). We further show via a Lyapunov analysis that the mapping of densities on potentials has chaotic regions in this case. In real space the situation is more subtle. V-representability is formally assured but the mapping is often chaotic making the actual construction of the potential almost impossible. The chaotic nature of the mapping, studied for the first time here, has serious consequences regarding the possibility of using TDDFT in real-time settings.

A fundamental conjecture lying at the very heart of quantum mechanical density functional theory (DFT) is the property of v-representability (VR) of a given density function, namely the existence a system of N particles that has this density in its quantum ground-state[1]. This, and a related issue, of "non-interacting" VR, have been explored in several important works[1-6]. The main result of these studies is that while VR in DFT is not guaranteed in general, even for reasonable looking densities, there exist rather flexible conditions for it (densities on lattices or non-degenerate ground states).

A similar issue arises in time-dependent (TD) DFT, when the "density on potential mapping" (DoPM) is considered. Runge and Gross[7] (RG), proved uniqueness, while Mearns and Kohn[8] found that VR cannot be guaranteed in temporal sinusoidal density fluctuations. This crack in TDVR was healed when Van Leeuwen considered "switch-on densities"[9] showing that VR is assured. A third aspect of the DoPM, largely neglected up to now is its chaotic nature. This issue is important in practical attempts to perform DoPM [10].

The present study explores VR, uniqueness and chaos in "switch-on" densities. First, we find that in finite lattice models TDVR is not always guaranteed, as shown already in the case of 2-level systems. Next, we consider a real-space system, represented on a Fourier grid. Here we find that although VR is formally assured, it is impossible to carry out DoPM, due to inherent instabilities. The consequences of these findings are discussed.

Let us start by considering VR in general following ref[9]. Suppose we are given a 3D particle density $n(\mathbf{r},t)$, $t \geq 0$ of $N_e$ particles and a compatible initial state $\psi_0$ such that the following two conditions hold:

$$n(\mathbf{r},0) = \langle \hat{n}(\mathbf{r}) \rangle_{t=0} \qquad \dot{n}(\mathbf{r},0) = \nabla \cdot \langle \hat{\mathbf{j}}(\mathbf{r}) \rangle_{t=0} \quad (1)$$

where $\hat{n}(\mathbf{r})$ and $\hat{\mathbf{j}}(\mathbf{r})$ are the operators of density and current density at $\mathbf{r}$ and $\langle \bullet \rangle_t \equiv \langle \psi(t) | \bullet | \psi(t) \rangle$. Denoting $\hat{F} = \hat{T} + \hat{U}$ as the sum of kinetic energy $\hat{T}$ and interaction energy $\hat{U} = \frac{1}{2} \sum_{i,j} v_{12}(|\mathbf{r}_i - \mathbf{r}_j|)$ of the particles we want to find the potential $v(\mathbf{r},t)$ for which the S.E

$$i\dot{\psi}(t) = \left[ \hat{F} + \int v(\mathbf{r},t) \hat{n}(\mathbf{r}) d^3r \right] \psi(t); \qquad \psi(0) = \psi_0 \quad (2)$$

yields a TD wavefunction $\psi(t)$ for which $n(\mathbf{r},t) = \langle \hat{n}(\mathbf{r}) \rangle_t$. From Heisenberg's equation of motion for the second time-derivative of the density, we find an implicit equation for $v$:

$$\hat{D}v(\mathbf{r},t) = -\left\langle \left[ \hat{F}, \left[ \hat{T}, \hat{n}(\mathbf{r}) \right] \right] \right\rangle_t - \ddot{n}(\mathbf{r},t) \quad (3)$$

Where, $\hat{D}$ is a linear "Hermitean" operator on potentials defined by: $\hat{D}v(\mathbf{r},t) = \int \left\langle \left[\hat{n}(\mathbf{r}'), \left[\hat{T}, \hat{n}(\mathbf{r})\right]\right] \right\rangle_t v(\mathbf{r}',t) d^3r'$. Van Leeuwen's showed that in real space under conditions that the density drops to zero at infinity $\hat{D}_{rs}v = -\nabla\cdot(n\nabla v)$. The integral form for $\hat{D}$ is useful in finite basis calculations. In real space and Fourier grid applications the operator is in fact essentially positive definite. Because of this latter property, VR is assured in TDDFT ($\hat{D}$ is invertible) and Eq. (3) combined with Eq. (2) gives the DoMP[9].

Before we analyze more fully the real space case, let us first consider a simple two level system with wavefunction $\psi = \begin{pmatrix} \psi_1 & \psi_2 \end{pmatrix}^T$. We now show that this system admits the RG theorem yet, we show an example where it is not VR. The Schrodinger equation is:

$$i\dot{\psi}(t) = \hat{H}(t)\psi; \quad \hat{H}(t) = V(t)\hat{\sigma}_z + \hat{\sigma}_x. \quad (4)$$

Where $\sigma_i$ ($i=x,y,z$) are the Pauli matrices:

$$\hat{\sigma}_x = \begin{pmatrix} 0 & 1 \\ 1 & 0 \end{pmatrix} \quad \hat{\sigma}_y = \begin{pmatrix} 0 & -i \\ i & 0 \end{pmatrix} \quad \hat{\sigma}_z = \begin{pmatrix} 1 & 0 \\ 0 & -1 \end{pmatrix}, \quad (5)$$

having the well-known cyclic commutation relations $2i\hat{\sigma}_x = [\hat{\sigma}_y, \hat{\sigma}_z]$. A term proportional to $\sigma_y$ was not included in the Hamiltonian, it corresponds to the presence of magnetic vector potentials.

The density is $\hat{n} = (\hat{\sigma}_z + 1)/2$ (clearly, knowing the density at one site is sufficient). The analog of the RG theorem in this case states that given a density $n(t)$ and a compatible initial state $\psi(0)$ obeying conditions analogous to Eq. (1):

$$n(0) \equiv \langle n \rangle_{t=0} \quad \dot{n}(0) \equiv \langle \sigma_y \rangle_{t=0}, \quad (6)$$

the potential $V(t)$ is uniquely determined. From Heisenberg's equations $\ddot{n}(t) = 2(V\hat{\sigma}_x - \hat{\sigma}_z)$ so the potential $V(t)$ affects $\ddot{n}(t)$ and the analog of Eq. (3) for this system is:

$$\sigma_x(t)V(t) = \ddot{n}(t)/2 + \sigma_z(t) \equiv f(t). \quad (7)$$

Clearly, $f(t) = \ddot{n}(t)/2 + 2n(t) - 1$, is determined solely by the given density. $V(t)$ is determined directly from Eqs. (7) and (4), obtaining the non-linear Schrödinger equation:

$$i\dot{\psi}(t) = \left(\frac{f(t)}{\sigma_x(t)}\hat{\sigma}_z + \hat{\sigma}_x\right)\psi(t) \quad (8)$$

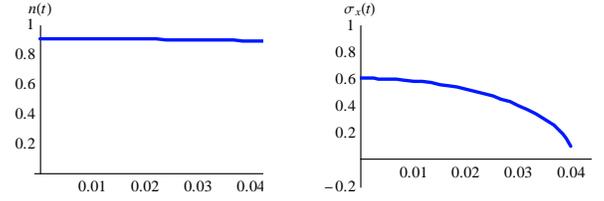

**FIGURE 1: THE DENSITY $n(t)$ (LEFT) AND $\sigma_x(t)$ FOR THE 2-LEVEL SYSTEM CASE STUDY.**

This is a direct analogy to the combination of Eqs. (3) and (2) in the real space system. Solving this equation yields the mapped potential: $V(t) = f(t)/\sigma_x(t)$. Indeed this potential is unique. Existence however is not assured because $\sigma_x(t)$ is not a positive definite. Thus at some $t_c$ the denominator may vanish while the numerator does not. This happens whenever the system localizes on one site. It also happens in other cases, namely when the relative phase of the two wave function components is such that $\text{Re}\,\psi_1^*\psi_2$ vanishes. Consider a numerical example. Take the density:

$$n(t) = (1 + \mu\cos\omega t)/2 \quad (9)$$

With $\mu = 0.8$ and $\omega = 2\pi$. The density has the property that $n(0) = 9/10$ and $\dot{n}(0) = 0$, thus a compatible initial state is the real state $\psi(0) = \begin{pmatrix} 3 & 1 \end{pmatrix}^T/\sqrt{10}$. The density has a period of $T = 1$ but we will only consider a small fraction of the first period, $t \in [0, 0.04]$. The density in this time interval is given in Figure 1 (Left), smooth and almost constant. The value of $f(t)$ is almost constant as well, equal to ~-0.7. Still, looking at $\sigma_x$, one sees that it quickly drops from an initial value of 0.6 to zero at $t_c \approx 0.041$. At this time the SE blows up and the potential becomes undefined. **Conclusion:** Here is a case where a naïve time-dependent density on a (2-site) lattice is non v-representable.

What about circumventing this, forcing an approximate solution on Eq. (7). A standard approach due to Tikhonov[11] is to replace $f/\sigma_x$ by $V = \sigma_x f/(\sigma_x^2 + \alpha)$, where $\alpha > 0$ is small, leading to the regularized Schrödinger Equation:

$$i\dot{\psi}(t) = \left(\frac{f(t)\sigma_x(t)}{\sigma_x(t)^2 + \alpha}\hat{\sigma}_z + \hat{\sigma}_x\right)\psi(t) \qquad (10)$$

This gives an approximate density $\tilde{n}(t)$. If the difference $\Delta n(t) = n(t) - \tilde{n}(t)$ is small, we can be satisfied with $\tilde{n}$ as an approximation. Sometimes this works. But it can also fail. In our case, a small value of $\alpha$ ($10^{-8}$) leads to insurmountable numerical instabilities, even when stiff solvers for Eq. (10) are used. Lowering the value of $\alpha$ to say $10^{-4}$ is leads to large divergence in precision as $t > t_c$ (Figure 2 (Left)).

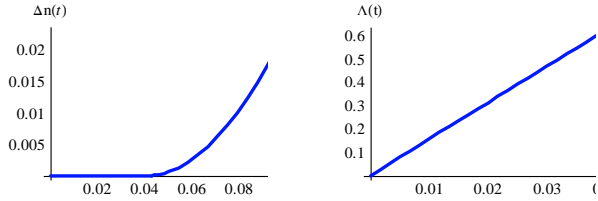

**FIGURE 2: (LEFT) THE ERROR IN DENSITY WHEN $\alpha = 10^{-4}$. (RIGHT) THE NON-EXPLOSIVE COEFFICIENT OF THE MAXIMAL LOCAL LYAPUNOV EXPONENT.**

The reason for the large deviances near and after $t > t_c$ are a result of instability of Eq. (8), as shown by a Lyapunov analysis[12], whose details will be described in a later publication. The wavefunction at time $t$ $\psi(t) = \xi + i\eta$ is decomposed into its real and imaginary parts. If a fluctuation $\delta\psi = \delta\xi + i\delta\eta$ is formed at this time it will evolve under the linearized equation $\left(|\delta\dot{\xi}\rangle \ |\delta\dot{\eta}\rangle\right)^T = \hat{L}\left(|\delta\xi\rangle \ |\delta\eta\rangle\right)^T$, where:

$$\hat{L} = \begin{pmatrix} 0 & \hat{H} \\ -\hat{H} & 0 \end{pmatrix} - \frac{f}{\sigma_x^2}\begin{pmatrix} \hat{\sigma}_z|\eta\rangle\langle\xi|\hat{\sigma}_x & \hat{\sigma}_z|\eta\rangle\langle\eta|\hat{\sigma}_x \\ -\hat{\sigma}_z|\xi\rangle\langle\xi|\hat{\sigma}_x & -\hat{\sigma}_z|\xi\rangle\langle\eta|\hat{\sigma}_x \end{pmatrix}. \qquad (11)$$

The stability of Eq. (8) is determined by the eigenvalue of $\hat{L}$ having the largest real part $\lambda(t) = \lambda' + i\lambda''$. If $\lambda' > 0$ then there will be fluctuations that grow in proportion to $e^{+\int_0^t \lambda'(t')dt'}$. $\lambda'(t)$ is called the local Lyapunov exponent[12]. Denoting the the local Lyapunov exponent for Eq. (8) by $\lambda(t) = -\Lambda(t)f(t)/\sigma_x^2$ we calculated numerically $\Lambda(t)$. This quantity, shown in Figure 2 (Right) is positive and grows more or less linearly with $t$. Since $f(t)$ is almost constant, with a value near $-0.7$, we find that that in the vicinity of $t_c$, where $\sigma_x(t)$ drops near zero the Lyapunov exponent $\lambda'(t)$ becomes huge and positive, rendering Eq. (8) highly chaotic near $t_c$: small fluctuations in the will cause an extremely large change in the wave function and the computed density will deviate wildly from the given density, i.e. $\Delta n$ will grow exponentially, as indeed it is seen Figure 2.

After considering the 2-level system, let us return to real-space, using the combination of Eqs. (3) and (2). The first problem one encounters is that in any typical quantum mechanical system there are regions where the density is very small (the asymptotes, for example) and this causes near singularities in $\hat{D}$ ($\hat{D}$ is positive definite except for a trivial singularity in that $\hat{D}v = 0$ whenever $v(\mathbf{r}) = const$, but this is easy to handle). These near-singularities lead to an ill-posed Eq. (3). Thus some sort of regularization is needed. We can use Tikhonov's regularization for the inverse of $\hat{D} = \sum_n d_n |w_n\rangle\langle w_n|$ (written in terms of its eigenvalues $d_n$ and eigenvectors $|w_n\rangle$) replacing $\hat{D}^{-1}$ by

$$\hat{D}^{-1} \to \sum_n i(d_n)|w_n\rangle\langle w_n|, \qquad (12)$$

where $i(d_n) = \frac{d_n}{d_n^2 + \alpha}$. $\alpha > 0$ is the small regularization parameter. In any application the value of $\alpha$ must be chosen carefully as a compromise between two Platonic ideals: precision and stability.

As in the 2-level system, the problem one encounters is that small errors (incurred by Tikhonov regularization) are sometimes amplified by the unstable nature of the nonlinear time propagation. We performed a stability analysis of the Lyapunov type here as well. A fluctuation in the wave function will grow in time approximately as $e^{\int_0^t \lambda'_n(\tau)d\tau}$ where $\lambda'_n$ is a the Lyapunov exponent. If the $\frac{1}{t}\int_0^t \lambda'_n(\tau)d\tau$ is positive, the solution is unstable, i.e. fluctuations will grow exponentially. We give now an explicit example where this happens (such an example was not difficult to find: practically any attempt to map a potential for a given density failed in a similar manner). Consider a 1-dimensional particle of unit mass in a Harmonic potential well $V(x) = \frac{1}{2}x^2$ with time-dependent density:

$$n(x,t) = \left(\cos\omega\tau(t)\psi_0(x) + \sin\omega\tau(t)\psi_1(x)\right)^2 \qquad (13)$$

Where $\psi_0(x)$ and $\psi_1(x)$ are the ground and first excited states of this potential. The function $\tau$ is not so important, except that we want to have $\dot{\tau}(0)=0$ so that $\psi_0(x)$ can be taken as the initial state. We chose: $\tau(t)=t^4/(1+t^3)$. We took $\omega=2\pi/100$ and considered only short times relative to the period (i.e. $t\ll 100$) so here too only relatively gentle perturbations of the density are considered. The density as a function of $x$ and $t$ for is shown in Figure 3 (Left).

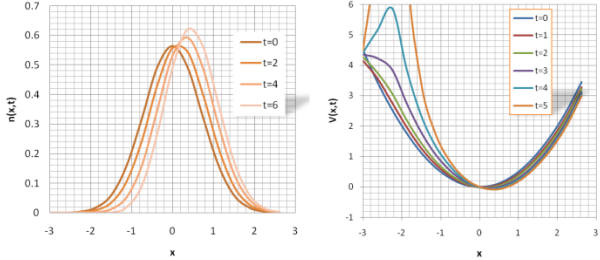

**FIGURE 3: (LEFT) THE DENSITY AS A FUNCTION OF TIME FOR THE MODEL PROBLEM. (RIGHT) THE CALCULATED MAPPED POTENTIAL.**

We solve the inversion equations, Eqs. (2) and (3), on a grid, spanning the range $x\in[-3,3]$ and using $N=16$ points. The representation is a based on the Fourier grid approach, employing periodic boundary conditions[13]. The non-linear time propagation was implemented using two different methods, which gave almost identical results, namely the explicit 5[th] order adaptive step size Runge Kutta (ERK) and a 3[rd] order implicit Runge Kutta method (IRK). The latter is suited for stiff problems. We took $\alpha=10^{-12}$ for regularization: eigenvalues of $\hat{D}$ much larger than $10^{-5}$ are treated almost exactly while those much smaller than $10^{-6}$ are almost completely neglected. This means that $\hat{D}$ has a condition number of about $10^6$ and therefore is not too susceptible to round-off errors (which are of the order of $10^{-13}$). The deviance between the given density $n(x,t)$ and $|\psi(x,t)|^2$,
$\Delta n \equiv \max_{i\in grid}\left|n_i(t)-|\psi_i(t)|^2\right|$ is shown in Figure 4. In the ERK calculation a moderate monotonic increase in the error is abruptly broken and a catastrophic increase is seen at $t_c=5.5$. This is accompanied by huge increase in calculation time since the time step of the adaptive algorithm quickly drops to a minute value due to the instability. In the IRK the algorithm does not explode but halts without converging to t=5.6. We have computed the eigenvalues of the Lyapunov operator $\hat{L}$ for this problem. These are given in Figure 4 (Right).

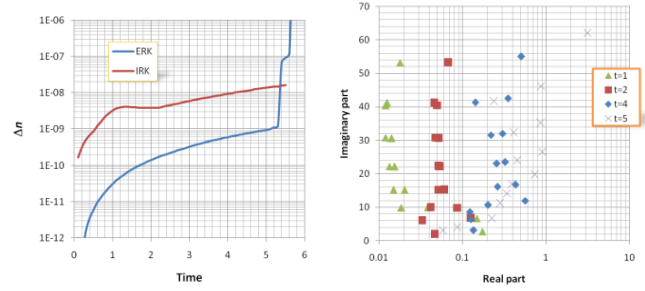

**FIGURE 4: (LEFT) THE DENSITY DEVIANCE $\Delta n(t)$. USING TWO NUMERICAL PROPAGATORS: IRK AND ERK. (RIGHT) THE LOCAL LYAPUNOV EXPONENTS (POSITIVE REAL PARTS) FOR SEVERAL TIMES.**

By computing Lyapunov exponents we have shown two cases where the mapping of densities on potentials is highly unstable, even in cases of "innocent looking" densities. The instabilities in real space are abundant in all our inversion attempts.

The consequence of the chaotic nature of the mapping is that it is extremely difficult (if not impossible) to actually find a potential that reconstructs a given time-dependent density. For TDDFT, the broader consequence here is that XC potentials $v_{XC}[n](\mathbf{r},t)$ are extremely sensitive functionals of the density $n$ and are thus ill-defined in strongly time-dependent problems. This severe conclusion can be relaxed somewhat by the following additional finding. We noticed that if the density at some point stops changing with time, the Lyapunov exponents quickly decrease. This issue is probably related to the great success of TDDFT within linear response.

We thank Dr. Y. Kurzweil for his help in preparing this paper.